\documentclass[a4paper,12pt]{article}

\usepackage{graphics}
\usepackage{latexsym}
\usepackage{amsmath}
\usepackage{amssymb}
\newcommand{\sectiono}[1]{\section{#1}\setcounter{equation}{0}}

\begin{document}
\begin{titlepage}
\thispagestyle{empty}
\begin{flushright}
UT-03-18\\
June, 2003 
\end{flushright}

\vskip 1.5 cm

\begin{center}
\noindent{\textbf{\LARGE{ Effective Gauge Degrees of Freedom \\\vspace{0.5cm}
 and the (Non)existence of 
\vspace{0.5cm}\\ the Glueball Superpotential
}}} 
\vskip 1.5cm
\noindent{\large{Yu Nakayama}\footnote{E-mail: nakayama@hep-th.phys.s.u-tokyo.ac.jp}}\\ 
\vspace{1cm}
\noindent{\small{\textit{Department of Physics, Faculty of Science, University of 
Tokyo}} \\ \vspace{2mm}
\small{\textit{Hongo 7-3-1, Bunkyo-ku, Tokyo 113-0033, Japan}}}
\end{center}
\vspace{1cm}
\begin{abstract}
We propose an efficient way to obtain a correct Veneziano-Yankielowicz type integration constant of the effective glueball superpotential $W_{eff}(S,g,\Lambda)$, even for massless theories. Applying our method, we show some $\mathcal{N} = 1$ theories do not have such an effective glueball superpotential, even though they have isolated vacua. In these cases, $S = 0$ typically.

\end{abstract}

\end{titlepage}
\newpage
\baselineskip 6mm


\sectiono{Introduction}\label{sec:introduction}
The chiral sector of $\mathcal{N}=1$ supersymmetric gauge theories has been studied for a long time (for a review see e.g.\cite{am} \cite{is}). Recently Dijkgraaf and Vafa conjectured a new method \cite{dv} to obtain the exact effective glueball superpotential using a matrix model technique. This conjecture was proved via integrating out all the colored freedom perturbatively in the presence of the gaugino condensation \cite{dglvz} \cite{iv}, or via using the (generalized) Konishi anomaly relations \cite{cdsw} \cite{csw}. The net result is, one can obtain (in the massive case) schematically
\begin{equation}
W_{eff}(S,g,\Lambda) = \text{Veneziano-Yankielowicz  term} + W_{pert}(S,g) ,
\end{equation}
where the Veneziano- Yankielowicz term is given as
\begin{equation}
\text{Veneziano-Yankielowicz term} = C(S,\Lambda) =-N_cS\left(\log{\frac{S}{\Lambda^3}} -1\right).
\end{equation}
This term can be interpreted as a measure factor of the matrix model calculation \cite{OV} \cite{cdsw}. 

In general, one can solve the (generalized) Konishi anomaly equations to determine $W_{eff} $ up to an ``integration constant factor" $C(S,\Lambda)$, which does not depend on other coupling constants: 
\begin{equation}
W_{eff}(S,g,\Lambda) = C(S,\Lambda) + W_{pert} (g,S).
\end{equation}
$W_{pert}$ can be obtained by the perturbative calculation (in special cases this reduces to the corresponding matrix model perturbative calculation), or by integrating the Konishi anomaly relations.

The problem is: how can we determine $C(S,\Lambda)$ then? It is independent of coupling constants so that we are able to take any convenient limit for its evaluation. For example, take the limit that all coupling constants except mass terms go to 0. In this limit, we should have precisely the Veneziano-Yankielowicz term, because at the low energy, the theory becomes the pure gluodynamics and the physics of the pure gluodynamics is captured by the Veneziano-Yankielowicz effective superpotential \cite{VY}. In principle, this completely determines the effective glueball superpotential (as long as one can introduce mass terms). In practice, however, the above-mentioned procedure is cumbersome when one is interested in the massless limit as we will see shortly. 

Think about the following example ($SU(N_c)$ SQCD with 1 flavor). We take the tree level superpotential as 
\begin{equation}
W_{eff} = m M + \lambda M^2,\label{eq:a} 
\end{equation}
where $ M = Q\tilde{Q}$. If we calculate $W_{eff}(S,m,\lambda,\Lambda)$ perturbatively around the $M=0$ vacuum, we will get
\begin{equation}
W_{eff} = C_1(S,\Lambda) + S\log{\frac{m}{\Lambda}} + a_1\lambda\frac{S^2}{m^2} +a_2\lambda^2\frac{S^3}{m^4} + \cdots.\label{eq:ex}
\end{equation}
In this case, taking $\lambda \to 0$ limit allows us to determine $C_1(S,\Lambda)$ as 
\begin{equation}
C_1(S,\Lambda) = -N_cS\left(\log{\frac{S}{\Lambda^3}} -1\right),
\end{equation}
which is just the original Veneziano-Yankielowicz term. However, suppose that we are interested in the massless limit. As is seen directly from the perturbation series (\ref{eq:ex}), $m \to 0$ limit is obtainable only after summing all the series above. 

Actually, in this case, one can sum up all the diagrams (or series) and the limit exist (Alternatively saying, one can integrate the Konishi anomaly relation exactly. See \cite{b} for details.). However, if we are interested in only $m=0$ case, integrating the Konishi anomaly relation is no more difficult than $m \neq 0 , \lambda =0$ case. The Konishi anomaly equation says\footnote{We use the same notation to the quantum mechanical gauge invariant operator and its vacuum expectation value, since they are almost the same thing in the sense of the chiral ring.},
\begin{equation}
 S = 2\lambda M^2, 
\end{equation}
which can be solved as
\begin{equation}
\frac{\partial W_{eff}}{\partial \lambda} = M^2 = \frac{S}{2\lambda}.
\end{equation}
We get right away 
\begin{equation}
W_{eff} = \frac{S}{2} \log{\lambda\Lambda} + C_2(S,\Lambda),
\end{equation}
where $C_2(S,\Lambda)$ is an integration constant which is to be determined. As a matter of fact, this corresponds to (\ref{eq:ex}) after taking $m\to 0$ limit.

How can we determine $C_2(S,\Lambda)$ ? One can show, taking massless limit of (\ref{eq:ex}),
\begin{equation}
C_2(S,\Lambda) = -\left(N_c-\frac{1}{2}\right) S\left(\log{\frac{S}{\Lambda^3}}-1\right) .\label{eq:d}
\end{equation}
Note that the coefficient of the Veneziano-Yankielowicz term is \textit{not} $N_c$. This mismatch appears because of the limiting procedure. It would be nice to gain this result without introducing a mass term and consequently without doing this kind of cumbersome limiting procedure. We propose how to obtain correct $C(S)$ efficiently in this paper.

Another example is a chiral model. As is chiral, one can not introduce a mass term to $W_{tree}$ by definition. Therefore, at first sight, it seems very difficult to determine $C(S)$. The clever analysis of \cite{b} tells us that one can use the Higgs mechanism instead, in order to effectively introduce mass terms and to match the result with the pure glue theory (whose color degrees of freedom are reduced by the Higgs mechanism.).

The natural question is whether this is applicable to any chiral theories. For this procedure to work, we should have pure glue theories after making all colored matters massive via the Higgs mechanism. It fails sometimes, however. For example, the Higgs mechanism may break the gauge group completely or down to $U(1)$. In this case, matching seems impossible. In particular, in $SU(2)$ theory, this is inevitable. Generally speaking, with more matters, more difficult it becomes to Higgs properly. There is also a technical problem. Even if we could Higgs chiral matters properly, we have to take an un-Higgsed limit if we are interested in un-Higgsed vacua. This requires just the same limiting procedure as was discussed above in the massive case. We have to sum up all the perturbation series to make any un-Higgsed limit sensible. Clearly, it will be a great advantage if one can tell the integration constant $C(S)$ without resorting to the Higgs mechanism. Our proposal applies to chiral theories just the same as to non-chiral theories.

The proposal is roughly as follows. We expand the effective superpotential:
\begin{equation}
W_{eff} = -C S \left(\log{\frac{S}{\Lambda^3}} -1\right) + a_1 S + a_2 S^2 + \cdots .
\end{equation}
We also expand $W_{tree}$ in $S$:
\begin{equation}
W_{tree} = b_1 S + b_2 S^2 + \cdots.
\end{equation}
Then we propose that we can determine $C$ via the following relation:
\begin{equation}
C- b_1 = N_c - \sum n_i,\label{eq:p}
\end{equation}
where $n_i$ is the Dynkin index of the $i$th matter chiral superfield. Let us check quickly how we can obtain $C(S)$ via this formula, in the example considered above (\ref{eq:a}) with $m=0$. In this case, $b_1 = \frac{1}{2}$ and $\sum n_i = 1$, which means, according to our proposal (\ref{eq:p}), $C = N_c-\frac{1}{2}$. This coincides with the direct limiting calculation (\ref{eq:d}).

So far, we have assumed the existence of the Dijkgraag-Vafa type effective glueball superpotential $W_{eff}(S,g,\Lambda)$. However, applying the above proposed formula, we can see that the existence of $W_{eff}(S,g,\Lambda)$ fails in some theories. In these theories, there exist some vacua which one can not obtain by extremizing $W_{eff}(S,g,\Lambda)$ by $S$. We will show some examples and analyze the general structure in this paper.

The organization of this paper is as follows. In section 2, we propose the efficient method to obtain the integration constant of the Konishi anomaly relations $C(S)$ and give its derivation based on the ILS (Intriligator-Leigh-Seiberg) linearity principle. We apply this method to the models which were studied in the literature and show its efficiency and correctness by comparing our results with their previous results. In section 3, applying our proposal to more exotic models, we study the case in which we do not have $W_{eff}(S,g,\Lambda)$. We find various examples including chiral theories and non-chiral theories. We see that there are some connections between the nonexistence of $W_{eff}(S,g,\Lambda)$ and the effective gauge degrees of freedom which are closely related with $C(S)$. In section 4, we summarize our results and discuss the possible further applications.

\sectiono{Effective gauge degrees of freedom}\label{sec:sd}

Our proposal is as follows. Consider $SU(N_c)$ gauge theory with any tree level superpotential which has discreet vacua classically. Use the Konishi anomaly relation \cite{k} or the generalized version of it \cite{cdsw} to determine the effective glueball superpotential up to a coupling independent ``integration constant" $C(S)$ as is discussed in section 1. Here by the Konishi anomaly relation, we mean
\begin{equation}
 \left\langle\frac{\partial W_{tree}}{\partial \phi_i} \phi_j\right\rangle = 2 n(\phi_i) S \delta_{ij},
\end{equation}
where $\phi$ is any matter chiral superfield and $n(\phi)$ is its Dynkin index\footnote{We take the normalization of the index as follows; for the fundamental representation $n_{fund} = 1/2$, for the adjoint representation $n_{adj} = N_c$ and so on.}. $S$ is the so called glueball superfield\footnote{If we have an adjoint matter, we need to define the glueball superfields  $S_i$ which represent the classically unbroken subgroups of the original gauge group. Though the calculation below readily applies to this case, we concentrate on the simple case for simplicity.}
\begin{equation}
S = -\frac{1}{32\pi^2}\mathrm{Tr} W^2 .
\end{equation}
We expand the effective glueball superpotential and the tree level superpotential (after substituting the Konishi anomaly relations) such that
\begin{equation}
W_{eff} = -C S \left(\log{\frac{S}{\Lambda^3}} -1\right) + a_1 S + a_2 S^2 + \cdots\label{eq:S1}
\end{equation}
\begin{equation}
W_{tree} = b_1 S + b_2 S^2 + \cdots\label{eq:S2}
\end{equation}
where we assumed that there are no singular terms other than Veneziano-Yankielowicz like one. Then we propose the following relation:
\begin{equation}
C- b_1 = N_c - \sum n_i
\label{eq:Pr}
\end{equation}
should be hold. Here $n_i$ is the Dynkin index of the $i$th matter chiral superfield.

This formula can be derived if one presupposes the ILS (Intriligator-Leigh-Seiberg) linearity principle \cite{ILS} and R charge conservation. The basis of this formula is the relation \cite{Tach};
\begin{equation}
W_{n.p.} \equiv W_{eff} - W_{tree} = (N_c - \sum n_i) S \label{eq:T}
\end{equation}
which was originally derived via the (massive) Dijkgraaf-Vafa conjecture in order to prove the ILS linearity principle. We are primarily interested in the massless case here, so we can not prove this relation from the first principle (at least not yet). Rather, we assume the ILS linearity principle that naturally leads to this relation as we will see.
Before using this relation, we would like to clarify its meaning here. If we assume the ILS linearity principle, that means $W_{n.p.}$ does not depend on any coupling constants (other than $\Lambda$). 
\begin{equation}
 W_{n.p.} = f(X) \Lambda^n 
\end{equation}
where $X$ represent all gauge invariant fields. $\Lambda$ is the dynamically generated mass scale and precisely it is defined via;
\begin{equation}
\Lambda ^b = e^{-\frac{8\pi^2}{g_0^2}} \Lambda_0^ b 
\end{equation}
where $\Lambda_0$ is the UV cutoff, $g_0$ is the bare gauge coupling and $b$ is the coefficient of the $\beta$ function, namely $b = 3N_c - \sum n_i$. The ILS linearity principle states that the effective superpotential is the sum of the dynamically generated $W_{n.p.}$ and the tree level superpotential $W_{tree}$ itself;
\begin{equation}
W_{eff} = W_{n.p.} + W_{tree} .
\end{equation}
From this one can calculate the gaugino condensation as follows (see e.g. \cite{ps})
\begin{equation}
 S = \frac{\partial W_{eff}}{\partial \Lambda^b} \Lambda^b = \frac{1}{N_c -\sum n_i} W_{n.p.} .
\end{equation}
The second equality holds because anomaly free R charge conservation enforces the power of $\Lambda^b$ in the effective superpotential to be $\frac{1}{N_c - \sum n_i}$ uniquely. This is what the relation (\ref{eq:T}) tells about.

Going back to the expansion (\ref{eq:S1}),(\ref{eq:S2}), at the first order in $S$, the extremization condition is 
\begin{equation}
\frac{\partial W_{eff}}{\partial S} = 0 \iff -C\log{\frac{S}{\Lambda^3}} +a_1 = 0.
\end{equation}
Substituting back to the effective superpotential, we get
\begin{equation}
W_{eff} = CS + O(S^2) .
\end{equation}

Also at the first order in $S$, we take as the tree level superpotential 
\begin{equation}
W_{tree} = b_1 S + O(S^2).
\end{equation}
Then we require (\ref{eq:T}) to be hold. It leads immediately to the relation 
\begin{equation}
C- b_1 = N_c - \sum n_i,
\end{equation}
which is just what we have proposed. We call this factor $C$ ``the effective gauge degrees of freedom" because of the reason we will see in the examples below.

Let us check the validity of this formula by applying to the models which were studied in \cite{b}. In the next section we will apply our method to more exotic models. 

The first example is $SU(N_c)$ SQCD with one flavor\footnote{There are other approaches to the massless SQCD both in the context of the Matrix model and the Konishi anomaly relation. See e.g. \cite{DJ} \cite{BRT} \cite{F} \cite{RTW} \cite{C}. See also \cite{Seiberg:2002jq} \cite{Argurio:2002xv} \cite{McGreevy:2002yg} \cite{Suzuki:2002gp} \cite{Bena:2002kw} \cite{Tachikawa:2002wk} \cite{Argurio:2002hk} \cite{Naculich:2002hr}  \cite{Feng:2002zb} \cite{Feng:2002yf} \cite{Ookouchi:2002be} \cite{Ohta:2002rd} \cite{Bena:2002tn} \cite{Hofman:2002bi} \cite{Demasure:2002jb} \cite{Suzuki:2002jc} \cite{Bena:2003vk} \cite{Feng:2003eg}.}. The tree level superpotential is
\begin{equation}
W_{tree} = \lambda M^2 + \alpha M^4.
\end{equation}
In this model, we can integrate the Konishi anomaly equation exactly and we get  \cite{Argurio:2002xv} \cite{b}
\begin{equation}
W_{eff} = C(S) -\frac{\lambda^2}{8\alpha}\mp\frac{\lambda^2}{8\alpha}\sqrt{1+\frac{4\alpha}{\lambda^2}S} + \frac{1}{2}S\log\frac{\alpha}{\lambda}-\frac{1}{2}S\log\left(\mp\sqrt{1+\frac{4\alpha}{\lambda^2}S}-1\right).
\end{equation}
Classically the upper sign corresponds to the Higgsed branch and the lower sign to the un-Higgsed one. Although this tree level superpotential does not have a mass term, we can unambiguously determine $C(S)$ in the Higgsed branch. As is discussed in \cite{b},
\begin{equation}
C(S) = - (N_c -1) S\left(\log{\frac{S}{\Lambda^3}} -1\right) .
\end{equation}
Intuitively we can interpret this that the gauge degrees of freedom which survive in the IR are reduced by 1 via the Higgs mechanism. What about the un-Higgsed branch? Let us see, using our method, we will reproduce the same answer also in this case.

First, we expand $W_{eff}$ in the series of $S$:
\begin{equation}
W_{eff} = -C S\left(\log{\frac{S}{\Lambda^3}}-1\right) -\frac{1}{2}S\left(\log{\frac{S}{\Lambda^3}}-1\right) +\frac{1}{2}S\log\lambda + O(S^2) .
\end{equation}
We also have in the linear order,
\begin{equation}
M^2 = \frac{S}{2\lambda} + O(S^2).
\end{equation}
Thus the tree level superpotential is 
\begin{equation}
W_{tree} = \lambda M^2 + \alpha M^4 = \frac{S}{2} +O(S^2).
\end{equation}
We apply our formula (\ref{eq:Pr}), 
\begin{equation}
\left(C+\frac{1}{2}\right) -\frac{1}{2} = N_c -1.
\end{equation}
Then we acquire the same conclusion $C = N_c - 1$, which is of course the correct result. Note that in the un-Higgsed branch, effective gauge degrees of freedom is $N_c-\frac{1}{2}$ as was calculated in section 1 in the $\alpha =0$ case (remember there is an extra factor $\frac{1}{2}$ from $\left[-\frac{1}{2}S\log\left(\mp\sqrt{1+\frac{4\alpha}{\lambda^2}S}-1\right)\right]$ in the un-Higgsed case).

The next example is a chiral $SU(6)$ model with 2 antifundamentals $\bar{Q}^I$ and 1 antisymmetric tensor $X$ \cite{MV} \cite{ADS1} \cite{ADS2} \cite{am} . The gauge invariant operators we will use are
\begin{equation}
 T = \epsilon_{IJ} \bar{Q}^I\bar{Q}^J X
\end{equation}
and
\begin{equation}
 U = \mathrm{Pf} X .
\end{equation}
where $I,J=1,2$ are the flavor indices and gauge indices are contracted in the obvious way. We consider the tree level superpotential
\begin{equation}
W_{tree} = hT + gU .
\end{equation}
In \cite{b}, introducing the auxiliary (unrenormalizable) interaction $\lambda TU$ and using the Higgs mechanism, they obtained as the effective glueball superpotential (in the $\lambda \to 0$ limit),
\begin{equation}
W_{eff} = -5S\left(\log{\frac{S}{\Lambda^3}-1}\right) + S \log{gh} .
\end{equation}
 
We reproduce this result without resorting to the Higgs mechanism (namely without introducing the auxiliary unrenormalizable interaction).

The Konishi anomaly relation states,
\begin{equation}
\begin{cases}
\; S = hT \\
\; 4S = hT +3gU 
\end{cases}  
\end{equation}
where we used the fact that the Dynkin index of an $SU(N)$ antisymmetric tensor is $\frac{N-2}{2}$.
We can solve these equations immediately,
\begin{equation}
\begin{cases}
\; T = \frac{S}{h} \\
\; U = \frac{S}{g} .
\end{cases}  
\end{equation}
Integrating this results,
\begin{equation}
W_{eff} = S\log{gh} + C(S) .
\end{equation}
It is also easy to see,
\begin{equation}
W_{tree} = 2S .
\end{equation}
Applying our proposal, 
\begin{eqnarray}
C-2&=&6-\left(2\cdot\frac{1}{2} + \frac{4}{2}\right) \cr
 \iff C&=&5 .
\end{eqnarray}
This is the desired result. Thus we can avoid the cumbersome limiting procedure and obtain the effective glueball superpotential directly.

\sectiono{Nonexistence of $W_{eff}(S,g,\Lambda)$}\label{sec:inexist}

We would like to apply our proposal and discuss the (non)existence of the effective glueball superpotential $W_{eff}(S,g,\lambda)$ in this section. To motivate, let us consider SQCD with $N_f$ flavors. We take as the tree level superpotential 
\begin{equation}
W_{tree} = g_k \mathrm{Tr} M^k,
\end{equation}
where $M_{ij} = Q_i\tilde{Q}_j $ is a meson superfield. We integrate the Konishi anomaly equation and obtain the effective glueball superpotential: 
\begin{equation}
W_{eff} = - C S\left(\log{\frac{S}{\Lambda^3}}-1\right) + \frac{N_f}{k} S\log{g_k\Lambda^{2k-3}}.
\end{equation}
We can determine $C$ just as was discussed in section 2.
\begin{equation}
 C = N_c - N_f + \frac{N_f}{k}.
\end{equation}
$C$ is the effective gauge degrees of freedom of this theory. What happens if $C = 0$ ? Take $k=2$ for example. To make $C = 0$, we should take $N_f = 2N_c$. This corresponds to the Seiberg self-dual point \cite{Sei}. 

Since $C=0$ in this case, we may conclude that there is no Veneziano-Yankielowicz term:
\begin{equation}
W_{eff} = \frac{N_f}{2} S\log{g_2 \Lambda} .
\end{equation}

If you demand the extremization of $S$:
\begin{equation}
\frac{\partial W_{eff}}{\partial S} = 0,
\end{equation}
then there is a contradiction unless
\begin{equation}
g_2 = \frac{1}{\Lambda} .
\end{equation}

This might remind you of the case of the IYIT (Izawa-Yanagida, Intriligator-Thomas) model \cite{IY} \cite{IT} which was also studied in this effective glueball superpotential approach in \cite{b}. In their analysis, they got the similar effective glueball superpotential which is just linear in $S$. Then they concluded that SUSY is dynamically broken in the IYIT model (except at some fine-tuned couplings).

Does our Seiberg self-dual theory considered above break SUSY dynamically? We can argue against this scenario in two ways.

First, it is believed (without the tree level superpotential) that the Seiberg dual point is in the conformal non-Abelian Coulomb phase. At the same time, anomaly free R charge of the operator $\mathrm{Tr} M^2 $ is 2. That means this tree level superpotential is a marginal perturbation. So we expect, after adding the tree level superpotential $g_2\mathrm{Tr} M^2 $, there still remains a conformal vacuum: $\langle S \rangle = \langle \mathrm{Tr} M^2 \rangle = 0$.

Another argument is as follows. Add $m \mathrm{Tr} M $ to the tree level superpotential, and one can show that there is a vacuum such that 
\begin{equation}
 S = O(m^2 \Lambda),
\end{equation}
which we hope becomes a conformal vacuum in the massless limit. However, we can not obtain this vacuum by extremizing an effective glueball superpotential $W_{eff} (S,g_2,\Lambda)$, even if we choose any $C(S)$ as an integration constant. If we begin with
\begin{equation}
W_{eff} = \frac{N_f}{2} S \log g_2 \Lambda + C(S),
\end{equation}
for $S=0$ to be the solution of the extremization equation,
\begin{equation}
\frac{N_f}{2} \log{g_2 \Lambda} = -C'(0)
\end{equation}
should be hold, whatever the value of $g_2$ is. However, since the right hand side does not depend on $g_2$, this is impossible.

Now, we will apply our method to more exotic models. The next example is $SU(7)$ with 6 antifundamentals $\bar{Q}_i$ and 2 antisymmetric tensors $A^I$. This model exhibits a dynamical SUSY breaking, if one adds an appropriate tree level superpotential \cite{dsb}. The relevant gauge invariant composite operators are
\begin{equation}
H_1 = A^1 \bar{Q}_1\bar{Q}_2, \ \ \ \ \   H_2 = A^2 \bar{Q}_2 \bar{Q}_3,
\end{equation}
\begin{equation}
H_3 = A^1 \bar{Q}_3\bar{Q}_4, \ \ \ \ \   H_4 = A^2 \bar{Q}_4 \bar{Q}_5,
\end{equation}
\begin{equation}
H_5 = A^1 \bar{Q}_5\bar{Q}_6, \ \ \ \ \   H_6 = A^2 \bar{Q}_6 \bar{Q}_1,
\end{equation}
and 
\begin{equation}
N_i = A^1A^1A^2A^2 \bar{Q}_i  \ \ \ \ \ ( i= 1,2,\cdots, 6) . 
\end{equation}

We first consider the case in which the tree level superpotential is
\begin{equation}
W_{tree} = f_1 H_1 +\cdots +f_6H_6 + g_1N_1 + \cdots +g_6N_6.
\end{equation}

Although we are ultimately interested in $g_i = 0$ case, we have to include these (unrenormalizable) terms to solve the Konishi anomaly equations as we will see shortly.

The Konishi anomaly equations are 
\begin{equation}
\begin{cases}
\; 5S = f_1H_1 + f_3 H_3 + f_5H_5 + 2\sum_ig_iN_i \\
\; 5S = f_2H_2 + f_4 H_4 + f_6H_6 + 2\sum_ig_iN_i\\
\; S = f_iH_i +f_{i+1}H_{i+1} + g_i N_i \ \ (\text{For} \ i = 1\cdots 6)\label{eq:sko},
\end{cases}
\end{equation}
in the third equation, no summation is implied. These equations are solved as
\begin{equation}
\begin{cases}
\; N_i = \frac{S}{3g_i} \\
\; H_i = \frac{S}{3f_i} .
\end{cases}
\end{equation}
Therefore, the effective glueball superpotential is 
\begin{equation} 
W_{eff} = C(S) + \frac{S}{3} \log(g_1\cdots g_6 \Lambda^{12}) + \frac{S}{3}\log(f_1\cdots f_6) \label{eq:sef}
\end{equation}
It is easy to see $W_{tree} = 4S$ and $\sum n_i = 2\cdot \frac{5}{2} + 6\cdot \frac{1}{2} = 8$, so that the effective gauge degrees of freedom $C$ is 3.
\begin{equation}
C(S) = -3S\left(\log{\frac{S}{\Lambda^3}}-1\right) .
\end{equation}
Extremizing the effective glueball superpotential determines the VEV of $N_i$s and $H_i$s\begin{equation}
\begin{cases}
\; N \propto \frac{f^{2/3}\Lambda^{13/3}}{g^{1/3}} \\
\; H \propto \frac{g^{2/3}\Lambda^{13/3}}{f^{1/3}} ,
\end{cases}
\end{equation}
where we have given all $f_i$s and $g_i$s the same value $f$ and $g$. From this result, we can conclude that, when $g \to 0$, the vacuum runs away and there remains no stable vacuum. This is compatible with the result discussed in the literature \cite{dsb}. This model ($g =0$) breaks SUSY dynamically.

Let us see whether we can obtain this conclusion, setting $g=0$ from the first. Setting $g=0$ in (\ref{eq:sko}), we immediately recognize that the Konishi anomaly equations allow an only solution $ S = H_i = 0$. This means that it is impossible to have an off-shell effective glueball superpotential $W_{eff}(S,f,\Lambda)$. Also one can see the same difficulty, when one takes $g\to 0$ limit in (\ref{eq:sef}). This tells you that if there were supersymmetric vacuum, this should be $S= H_i = 0$. How can we tell whether there is such a vacuum or not? The nonexistence of $W_{eff}(S,f,\Lambda)$ in the $g\to 0$ limit seems to suggest that there is not such a vacuum. However, this is not at all conclusive! In the following, we show a counterexample in which there \textit{is} such a vacuum but there does \textit{not} exist $W_{eff}(S,g,\Lambda)$ in the limit.

Consider the same model but with the tree level superpotential
\begin{equation}
W_{tree} = h_1 (H_1)^2 +\cdots +h_6(H_6)^2 + g_1N_1 + \cdots g_6N_6.
\end{equation}
The calculation is almost the same with the above one. The Konishi anomaly equations are 
\begin{equation}
\begin{cases}
\; 5S = 2h_1(H_1)^2 + 2h_3 (H_3)^2 + 2h_5(H_5)^2 + 2\sum_ig_iN_i \\
\; 5S = 2h_2(H_2)^2 + 2h_4 (H_4)^2 + 2h_6(H_6)^2 + 2\sum_ig_iN_i\\
\; S = 2h_i(H_i)^2 +2h_{i+1}(H_{i+1})^2 + g_i N_i  \ \ (\text{For} \ i = 1\cdots 6)\label{eq:sko2}.
\end{cases}
\end{equation}
These equations are solved as
\begin{equation}
\begin{cases}
\; N_i = \frac{S}{3g_i} \\
\; (H_i)^2 = \frac{S}{6h_i} .
\end{cases}
\end{equation}
Therefore, the effective glueball superpotential is 
\begin{equation} 
W_{eff} = C(S) + \frac{S}{3} \log(g_1\cdots g_6 \Lambda^{12}) + \frac{S}{6}\log(h_1\cdots h_6\Lambda^{18}) \label{eq:sef2}
\end{equation}
We can show that the effective gauge degrees of freedom $C = 2$ in this case. After extremizing the effective glueball superpotential, we obtain
\begin{equation}
\begin{cases}
\; N \propto h^{1/2}\Lambda^{13/2}\\
\; H \propto \frac{g\Lambda^{13/2}}{h^{1/2}} .
\end{cases}
\end{equation}
This shows that there exists a vacuum in the limit $g\to 0$. There, one can see $S=0$. However, if we take $g\to 0$ in (\ref{eq:sko2}) or in (\ref{eq:sef2}), we may misleadingly conclude that there does not exist $W_{eff}(S,h,\Lambda)$, so dynamical SUSY breaking occurs.

The last example we would like to discuss is the ISS (Intriligator-Seiberg-Shenker) model \cite{ISS}. This model is a $SU(2)$ gauge theory with one spin $\frac{3}{2}$ representation which we call $\psi$. The tree level superpotential we consider\footnote{Note, one can not make any gauge invariant bilinear in this model. That means this is a chiral theory.} is 
\begin{equation}
W_{tree} = \lambda u,
\end{equation}
where $ u = \psi^4$. The Konishi anomaly equation becomes 
\begin{equation}
 10 S = 4\lambda u .
\end{equation}
Solving this and substituting back into the tree level potential gives
\begin{equation}
W_{tree} = \frac{5}{2} S .
\end{equation}
We consider the effective glueball superpotential;
\begin{equation}
W_{eff} = -CS\left(\log{\frac{S}{\Lambda^3}}-1\right)+\frac{5}{2} S\log{\lambda \Lambda} 
\end{equation}
In this case, our formula (\ref{eq:Pr}) determines $C$ as 
\begin{equation*}
 C - \frac{5}{2} = 2 - 5 
\end{equation*}
\begin{equation}
C = -\frac{1}{2} .
\end{equation}

Thus as long as we believe our procedure is correct, the final effective glueball superpotential is
\begin{equation}
W_{eff} = \frac{1}{2}S\left(\log{\frac{S}{\Lambda^3}}-1\right)+\frac{5}{2} S\log{\lambda \Lambda} .
\end{equation}
By extremizing $W_{eff}$ by $S$, we obtain
\begin{equation}
 S = \frac{1}{\lambda^5\Lambda^2}
\end{equation}
\begin{equation}
 u = \frac{5}{2\lambda^6\Lambda^2}.
\end{equation}

Is this result correct? This vacuum implies that there is a dynamically generated nonperturbative superpotential
\begin{equation}
W_{n.p.} = a\frac{u^{\frac{5}{6}}}{\Lambda^{\frac{1}{3}}}.
\end{equation}
Nevertheless, this superpotential was discarded in the literature \cite{ISS}, because it has no sensible behavior in the classical limit. 

Since we have obtained an unwanted vacuum, we can not say anything about whether this model breaks SUSY dynamically or not. Actually the very problem discussed in the literature \cite{ISS} is whether this theory has an $S=u=0$ vacuum. We are forced to feel that something is wrong with our approach. One of the manifestation of this peculiarity is that, in this model, the effective gauge degrees of freedom $C$ becomes negative, which seems to be hard to interpret.

Let us review our logic carefully. We assumed the existence of $W_{eff}(S,g,\Lambda)$. Under the assumption of its existence (and the ILS linearity principle), we can use our method to determine it. Thus, if there is not such a thing (because of the dynamical SUSY breakdown or some other reasons), we can not obtain anything (or might obtain, in principle, anything that are wrong as the effective glueball superpotential). Also, as is discussed in the chiral $SU(7)$ case, the nonexistence of $W_{eff}(S,g,\Lambda)$ does not necessarily imply the dynamical SUSY breakdown. This is the way things are. We should invent another method to see whether the ISS model breaks SUSY dynamically\footnote{The direct way to see the possibility of the $S=0$ vacuum is to study the Kahler potential of the theory. However, this is beyond our current ability.}.

Before concluding this section, we would like to discuss the nature of the missing vacua and the effective gauge degrees of freedom. In all the examples considered above, missing vacua correspond to $S = 0$ conformal phase\footnote{This reminds us of the Kovner-Shifman vacuum \cite{Kovner:1997im}.}. The reason seems clear intuitively, because in the conformal phase, there is no gaugino condensation and the vacuum structure is completely different from that of the confining or Higgsed vacua\footnote{Note that the confining vacua and the Higgsed vacua are on the other side of the same coin, see \cite{cc} for example.}, and it is natural to expect that the method suitable for finding the confinement vacua fails in this case. Then, there seems to be a natural connection between the effective gauge degrees of freedom $C$ and the existence of the missing vacua, although this is not always the case. Roughly speaking, when $C \le 0$, remaining gauge degrees of freedom do not confine and the conformal phase can be possible. We conjecture this is one of the mechanisms which forbid us to obtain the effective glueball superpotential $W_{eff} (S,g,\Lambda)$\footnote{Another possibility is that we do have the right superpotential, but the Kahler potential of $S$ becomes singular at $S=0$. This is plausible, since we expect nontrivially interacting gluinos and gluons there. In this case, the actual effective potential of $S$ is $ V_{eff}(S,\bar{S}) = g^{S\bar{S}}\left|\frac{\partial W_{eff}}{\partial S}\right|$, so that it can become 0, provided $g^{S\bar{S}}(S,\bar{S}) =0$.}.

\sectiono{Summary and Discussion}\label{sec:summary}

We have shown in the above sections, under the assumption $W_{eff} (S,g,\Lambda)$ exits, how to efficiently obtain the Veneziano-Yankielowicz type part in the effective glueball superpotential $C(S,\Lambda)$, which is coupling independent. We used the ILS linearity principle as a guide. Therefore, if one could prove our proposal directly (for example, by the careful study of the matrix model measure and the suitable massless limiting procedure), it would mean the proof of the ILS linearity principle with any tree level superpotential without mass terms. It is a challenging problem.

As for the existence of the Dijkgraaf-Vafa type effective superpotential $W_{eff}(S,g,\Lambda)$, we have given some examples in which we do not have such an effective glueball superpotential or even though it has one, this effective glueball superpotential lacks some pieces of information about possible vacua. Of course, if one introduces a mass term, this possibility is ruled out. Conceptually it is clear: when the fields are all massive, integrating out all the colored matter makes sense and the well-defined glueball superpotential will exist. Therefore, for a non-chiral theory, the solution of this problem is clear in principle. We should consider the most general tree level superpotential including mass terms and take the desired limit carefully, though this is just what we have wanted to avoid in this paper\footnote{In this perspective, the nonexistence of $W(S,g,\Lambda)$ is formally clear. Once one allows all the couplings, one expects the effective superpotential depends on them ``holomorphically" (including logarithmic or cut singularities). Since everywhere nonsingular holomorphic function must be a constant, the effective superpotential should have some singularities in this enlarged space of coupling constants. This is just the reason why we can not obtain the effective glueball superpotential at some special couplings. However, when and where this happens or the nature of its singularity (conformal vacua, dynamical SUSY breakdown $\cdots$) are interesting questions to ask. Remember, as we saw in the introduction, in SQCD with one flavor, the finite order perturbative calculation suggests a singularity when we take $m \to 0$. Actually, $m\to0$ is \textit{not} singular at all. In contrast, when $N_f=2N_c$, there \textit{is} a true singularity there.}. For a chiral theory, the Higgs mechanism may work sometimes, but the general solution seems to be lacking. We also conjecture that these missing vacua always do not have the gaugino condensation i.e. $S=0$. This can also be seen in the Feynman diagrammatic approach of \cite{dglvz}, where $S\neq 0$ means an effective (fermionic) mass of the propagator, which can become singular in the ``massless" limit $S\to 0$.

We conclude this paper by suggesting further applications of our proposal.
\begin{itemize}
	\item Obviously we can apply our method to the study of the higher critical tree level superpotential. For example, in \cite{es} new $\mathcal{N}=1$ AD points, whose tree level superpotential requires a higher critical interaction, were discovered\footnote{Unfortunately, it can be shown in this case that the direct use of our method fails to capture the $\mathcal{N}=1$ AD vacuum (though it should be there, if one includes all the possible tree level couplings and takes the limiting procedure properly). This might be another example in which $W_{eff}(S,g,\Lambda)$ does not exist.}.
	\item Another arena we can apply our method is chiral models. In this paper, we have limited ourselves to the simpler models, where the Konishi anomaly relations are easily solved and we do not need to use the generalized version. Since our method does not depend on how the coupling dependent part is solved, it would be nice to apply it to more complex models in which more interesting physics might be found.
	\item We can use our formula that determines the effective gauge degrees of freedom in order to search interesting theories. We conjecture that, when the effective gauge degrees of freedom becomes zero or negative, something peculiar may happen. One possibility is the dynamical SUSY breakdown and another one is the nonexistence of the effective glueball superpotential and the consequent emergence of the (possibly conformal) vacuum in which the gaugino condensation is 0. We might encounter other various scenarios that we have not expected.
\end{itemize}


\section*{Acknowledgements}
The author would like to thank T. Eguchi, R. Nobuyama and Y. Tachikawa for very useful discussions.
\nopagebreak


\end{document}